\documentclass[12pt]{article}
\usepackage{epsfig}

\parskip=\medskipamount % space between paragraphs (TeX)
plus.3em minus.5em \abovedisplayshortskip=.5em plus.2em minus.4em
\renewcommand{\thesection}{\arabic{section}}

\catcode`@=11

\@addtoreset{equation}{section} \@addtoreset{equation}{subsection}
\def\theequation{\ifnum\value{section}=0 \arabic{equation}\ignorespaces
\else \ifnum\value{section}=-1 A.\arabic{equation}\ignorespaces
\else \ifnum\value{subsection}=0
\thesection.\arabic{equation}\ignorespaces \else
\thesection.\arabic{subsection}.\arabic{equation}\ignorespaces
                             \fi
                        \fi
                   \fi}

{\catcode`\'=\active \def'{{}^\bgroup\prim@s}}

\catcode`@=12

\newcommand{\bq}{\begin{equation}}
\newcommand{\be}{\begin{equation}}
\newcommand{\fq}{\end{equation}}
\newcommand{\ee}{\end{equation}}
\newcommand{\bqr}{\begin{eqnarray}}
\newcommand{\beqs}{\begin{eqnarray}}
\newcommand{\fqr}{\end{eqnarray}}
\newcommand{\eeqs}{\end{eqnarray}}

\def\bop#1{\setbox0=\hbox{$#1M$}\mkern1.5mu
    \vbox{\hrule height0pt depth.04\ht0
    \hbox{\vrule width.04\ht0 height.9\ht0 \kern.9\ht0
    \vrule width.04\ht0}\hrule height.04\ht0}\mkern1.5mu}
\begin{document}
\thispagestyle{empty}

\begin{flushright}
\begin{tabular}{l}
% TEP- \\
\end{tabular}
\end{flushright}

\vskip .6in
\begin{center}

{\Large\bf  A Novel Data Compression}

\vskip .6in

{\bf Gordon Chalmers}
\\[5mm]
% {\em address \\
%      address } \\

{e-mail: gordon@quartz.shango.com}

\vskip .5in minus .2in

{\bf Abstract}

\end{center}  

A novel data compression scheme is presented.  The method is very suitable 
for black and white images, and it can generate a compression factor 
of eight; in general the bitmap is optimized for an arbitary number of 
colors and not only for unused information contained in a conventional 
bitmap with $2^j$ bits per pixel.  
This compression method can be incorporated with other compression 
methods; it is suitable for other image types and audio.  The potential 
high compression factor is cost effective for both memory 
and bandwidth requirements.

\setcounter{page}{0}
\newpage
\setcounter{footnote}{0}

Data compression is increasingly important due to the size and type of 
information required to be manipulated.   There are many ways of compressing 
data.  An often overlooked aspect is to eliminate the redundancy of disk 
space within the actual byte.  Each byte consists of eight digits, and an 
image sometimes only requires one entry, which is a redundancy of an order 
of magnitude; in a colored grayscale, an optimistic factor of two in 
compression is beneficial.  Eliminating this redundancy is quite simple.   

Examine a black and white image with 768x768 pixels when each pixel is  
associated with a byte.  The bit can take values 1 or 0 when the image is 
monotone, but can range in more values when associated with color or a gray 
scale.  In the color example, there are 256 values contained in the byte, 
with a fraction of the values unused.  This algorithm eliminates the 
wasteful unused fraction.  

The entire image could be associated with a single number using a polytopic 
definition of the image \cite{ChalmersPoly}.  As an example, consider one 
row of the image, and specify the image by a base expansion, 

\bqr 
N=\sum_{i=0}^{768} a_i 2^i \ .  
\label{imageexpbasetwo} 
\fqr 
The base reduction of the number specifies the line's image as entries of 
$a_i$, either one or zero.  In a colored context with $N_c$ colors, the 
number could be expanded as 

\bqr 
N=\sum_{i=0}^{768} b_i N_c^i  \ , 
\label{imageexpbaseNc}
\fqr 
with the the numbers $b_i$ ranging from $0$ to $N_c-1$.  Subsequently, 
the number is stored in the usual fashion in the computer's memory, 
using the required number of bytes. 

Each individiual number specifies the line's image, but in grouping the 
image as a number, and in a colored base,  the redundancy of the byte is 
eliminated.  In a black and white format, the eight entries of the byte 
are now associated with eight pixels.  This compression factor is non-trivial, 
but quite simple to implement.  

The number $2^{769}$ requred to store the line's information seems large, 
as it has 231 digits in base ten.  However, the data storage required in 
the actual memory requires only 769 entries, and is only 96 bytes in total.  
The compression of a factor eight also has no loss of information.   In 
the case of the colored context, the compression ratio is still non-trivial, 
but not quite an order of magnitude; the number of colors versus the bytes 
per pixel is the limiting factor.

The data compression is also suitable to implement with other compression 
schemes, such as the simple RLE approach.  Count the consecutive $1$'s or 
$0$'s and insterpace a number with a marker to label the continuity.  The 
marked number could be an 8-bit number immediately adjacent to the occurance 
of the $0$ and placed within the number $N$.  The redundancy of the bytes 
is still eliminated with the $1$ or $0$ specifying the pixel reduced to a 
single bit, and not a byte, or a fraction of a byte in the 'colored' example.  

This compression scheme is quite simple to implement, but it appears not 
be in the literature.  A compression factor of eight is surely not to be 
overlooked, as hardware associated with memory allocation is quite costly 
when large amounts of data are required.  Also, the simplest image examples,   
which are black and white images, can have an impact on data 
communication; a reduction of almost an order of magnitude in bandwidth can 
be quite appealing.  In the colored example, the unused values of the 
pixel memory are eliminated; for example, using an 8-bit per pixel 
memory allocation for only an odd number such as 50 'colors' would lead 
to a further double in the compression. 

The compression scheme described here is very conservative, and has no 
loss of information.  It is suitable in a wide range of contexts.  Also, 
the polytopic definition of the image as a number can possibly examined 
in further compression schemes \cite{ChalmersUnpublished} involving 
only information as a pure number.

\vfill\break

\end{document}